\documentstyle[aps,prb,multicol,floats,epsfig,epsf]{revtex}
\begin{document} 
\draft
\wideabs{

\title{Topological complexity, contact order and protein folding
rates}
\author {P.F.N. Faisca$^{1,2}$ and R.C. Ball$^{1}$ }
\address{$^{1}$Department of Physics, University of Warwick, Coventry
CV4 7AL, U.K.}
\address{$^{2}$CFTC, Av. Prof. Gama Pinto 2, 1649-003 Lisboa Codex,
Portugal}
\maketitle
\begin{abstract}
Monte Carlo simulations of protein folding show the emergence
of a strong correlation between the relative contact order parameter, $CO$,
and the folding time, $t$, of two-state folding proteins for longer chains with
number of amino acids, $N \geq 54$, and higher contact order, $CO > 0.17$. 
The correlation is particularly strong for $N=80$ 
corresponding to slow and more complex folding kinetics.
These results are qualitatively compatible with experimental
data where a general trend towards increasing $t$
with $CO$ is indeed observed in a set of proteins with chain length
ranging from 41 to 154 amino acids.
\end{abstract}
}

\section{INTRODUCTION}
The search for correlations between the protein folding kinetics
and the native state equilibrium properties (i.e. chain length and
stability) presents a major challenge for those working in the field
of protein folding, both in theory and experiments. Progress has been
significantly hindered by difficulty analyzing the folding of protein
molecules larger than about 100 amino acids, whose kinetics is widely believed to
be based on some multiexponential mechanism {\cite{ALLAN}}. By contrast, for smaller
proteins whose folding kinetics is close to single exponential, there
seems to be some consensus as to the
dependence of the folding time, $t$, on native state stability
\cite{PLAXCO1,PLAXCO,THOMAS,PFN}.\par 
Experiment and theory
appear to be at odds with each other over the dependence of the 
folding time on the number $N$ of amino acids in the folding unit.
Recent Monte Carlo (MC) simulations {\cite{PFN}} of a simple lattice model
have proposed that, for two-state proteins, a scaling law
of the type $t\approx N^{\lambda}$, $\lambda \approx 5$, appropriately
describes the dependence of the folding
time on the chain length, $N$; a weaker dependence
($\lambda \approx 4$) has been previously reported
in Ref.6 for the same model Hamiltonian and distribution
of contact energies, and in Ref.7 for a two-letter alphabet
model that, apart from the
commonly used isotropic contact interactions,
also considers orientation-dependent
interactions. However, available experimental data
shows no correlation between $t$ and $N$
\cite{PLAXCO1,PLAXCO,DEMCHENKO}. 

\par  We examine here the influence of the native state geometric
properties on the protein folding kinetics in the context of MC simulation.
One simple parameter of the geometry which has already attracted
attention is contact order, measuring the average length of the backbone
loops connecting contacting pairs of residues in the
structure {\cite{PLAXCO1}}. Formally, the relative contact order,
$CO$, is defined as 
\begin{equation}
CO=\frac{1}{LN}\sum_{i,j}^N \Delta_{i,j}\vert i-j \vert,
\end{equation}    
where $N$ is the total number of amino acid residues in the protein, 
$L$ is the total number of contacts, and $\Delta_{i,j}=1$ if residues $i$ and $j$ are
in contact and is 0 otherwise. $\vert i -j \vert$ is the
backbone separation between residues $i$ and $j$.
High values of $CO$ are associated with protein structures where amino
acid residues interact on average with others that are far away in sequence
(long-range interactions), while those displaying predominantly local 
interactions are of low contact order.\par
A high correlation was found between the $CO$
parameter, and the folding rates for the protein set considered in
Ref.3: 
proteins with 'low' contact order tend to fold faster
than proteins with 'high' contact order.
This finding strongly supports the view that
native geometry strongly influences the kinetics of the 
rate-limiting step in the two-state mechanism of small
protein molecules ($N < 100$), determining their folding rates.\par
The connection between $CO$ and the dominant range of residue
interactions brings back the controversial issue of the importance of
local (and non-local) contacts in the dynamics of protein folding.
An argument against local contacts is that 
they might increase the 'roughness' of the energy
landscape, and therefore the stability of the unfolded state
{\cite{ABKEVICH}}. On the other hand, an argument supporting local
interactions is based on the idea that
they might provide the ideal substrate for the development of
nucleation or initiation sites, small local sequence substructures forming
in an early stage of folding and driving the subsequent pathway
{\cite{Wetlaufer}}. Moreover, the formation of non-local contacts in
early folding is entropically costly as it restricts the number of
conformations available to the folding unit {\cite{BAKER}}.\par
The limited amount of experimental information available
\cite{PLAXCO1,PLAXCO,LINDBERG} suggests that investigating this problem within the
scope of theoretical models could give more insight.

\section{MODEL AND METHODS}
To achieve this goal we consider a simple three dimensional lattice
model of a protein molecule whose Hamiltonian is given by the contact
approximation,
\begin{equation}
H(\lbrace \sigma_{i} \rbrace,\lbrace \vec{r_{i}} \rbrace)=\sum_{i>j}^N
\epsilon(\sigma_{i},\sigma_{j})\Delta(\vec{r_{i}}-\vec{r_{j}}),
\label{eq:no1}
\end{equation}
where $\lbrace \sigma_{i} \rbrace$ stands for an amino acid sequence
($\sigma_{i}$ being the chemical identity of bead $i$) while $\lbrace
\vec{r_{i}} \rbrace$ is the set of bead coordinates that define
a certain conformation. The contact function, $\Delta$, equals $1$ if
beads $i$ and $j$ are in contact but not covalently linked, and is $0$
otherwise. We follow many previous studies in taking the interaction
parameters $\epsilon$ from the $20 \times 20$
Miyazawa-Jerningan matrix derived from the distribution of contacts
in native proteins {\cite{MJ}}. 
Our folding simulations follow the standard MC Metropolis algorithm {\cite{METROPOLIS}}
and the kink-jump MC move set (end-move, corner-flip, and crankshaft){\cite{BINDER}}. 

\vspace{0.5cm}
\begin{table*}
\begin{tabular}[b]{c c c} 
\hspace{0.5cm}$N$ \hspace{0.5cm}  &  \hspace{0.5cm}$<E>$
\hspace{0.5cm}
&\hspace{0.5cm} $\sigma$ \hspace{0.5cm}\\ \hline
$36$ & $-15.8722$ & $0.0194$  \\
$48$ & $-23.1407$ & $0.0331$ \\
$54$ & $-28.8028$ & $0.0158$ \\
$64$ & $-35.0811$ & $0.0477$ \\
$80$ & $-46.2858$ & $0.0375$ 
\label{tab:tab1}
\end{tabular}
\medskip
\caption{Average sequence energy under the training scheme
of Shakhnovich and Gutin. For each chain length, $N$, the average is computed
over the total number of sequences ($\approx 2000$) designed for each set of targets.
$\sigma$ is the standard deviation of the mean.}
\end{table*}

\par

\section{NUMERICAL RESULTS}
\subsection{Contact order and homopolymer kinetics \label{sec:3a}}
We have explored the distribution of the relative contact order parameter
over a population of 2000 maximally compact target geometries found by
homopolymer relaxation \cite{GUTIN}.
This distribution is shown in Figures~1(a)-(e) for each of the studied
chain lengths $N=36$,$48$,$54$,$64$ and $80$, which are all
commensurate with folding to fill a simple cuboid. There is only a
slight shift in the modal contact order with chain length, from $0.19$
for $N=36$,$48$ down to $0.17$ for higher $N$.
However, the target fraction in the histogram tail
($CO\geq 0.22$) is significantly smaller for $N\geq 54$ 
than for shorter chains. Interestingly, the values of $CO$ found for
these lattice proteins
span approximately the same range as those found in real kinetically
characterized single domain proteins ($0.0745 \leq CO \leq 0.2120$)
{\cite {PLAXCO}}.\par  
We have also checked the intrinsic kinetic accessibility of the
compact configurations obtained,  by measuring the time $t_{col}$ for
these configurations to be reached under homopolymer relaxation.
Figure~1(f) shows there is no evident correlation between  $t_{col}$ and CO .\par 
\subsection{Finding the optimal folding temperature} 
To investigate the relationship between protein folding and
contact order, (at least) 20 target conformations for
each chain length, were selected so as to sample uniformly across
the range of contact order.  For each target an ensemble of 100 designed
sequences was prepared by using the design method developed
by Shakhnovich and Gutin {\cite {SG}} 
based on random heteropolymer theory, and simulated annealing
techniques. 
The average trained sequence energy, $<E>$, is shown in
Table I along with the standard deviation of the energy
distribution, $\sigma$.
Except for $N=54$, the chemical composition of the designed sequences
was the same as the one used in
Ref.5. In that study it was shown that
the optimal folding temperature, $T_{fold}(N)$, defined as the temperature
that minimizes the folding time, is close to a self-averaging
parameter.\par
Since the Shakhnovich and Gutin design scheme
preserves the overall sequence chemical composition we can safely
use for 36 and 48 bead long sequences studied here 
the $T_{fold}(36)$ and $T_{fold}(48)$ found in Ref.5.\par
For the longer $N$, and most particularly $N=80$,
foldicity, defined as the fraction of successful folding runs over
the total number of attempted runs, was for the vast
majority of the targets less than unity. This forced us to define the
optimal folding temperature, $T_{fold}(N)$, is such cases as the temperature
which optimized foldicity rather than the temperature which minimized the
folding time. The case $N=48$ sits at the margin and provides confirmation
that the two approaches to $T_{fold}(N)$ are not in conflict, as shown in
Figure.~2.\par

\subsection{Contact order and folding kinetics}
After determining $T_{fold}(N)$ we ran a MC folding 
simulation for every designed sequence. The simulations proceeded
until $\tau_{max}(N)$ MC steps or until folding was observed. 
The value of $\tau_{max}(N)$ was
chosen such that it was much longer than the typical 
folding time of the studied sequences.\par 
Fig.~3 shows the dependence of the folding time, $t$, on
the contact order parameter for chain lengths $N=36$ and $N=48$. The
folding time was computed as the mean first passage time averaged over 100
simulation runs. In either case the points are close to be uniformly
distributed suggesting no correlation between the $CO$, and the folding
time for these chain lengths. 
Here and elsewhere error bars indicate $\pm$ one standard error in the
mean.

\par
Figures~4(a)-(c) show the dependence of foldicity on
$CO$ for $N=54,64$ and $N=80$ respectively.
The results presented in Figures.~4(d)-(f) show, for the same chain lengths,
the dependence of the estimated folding time on the relative 
contact order paramenter. Two distinct scenarios emerge from the analysis
of the graphs:
\begin{enumerate}
\item{For $CO \leq 0.17$ there is no correlation between foldicity
(or folding time) and the relative contact order parameter;}
\item{For $CO>0.17$, a general trend towards decreasing foldicity with
increasing relative contact order can be observed. In this regime, 
a considerably strong positive correlation of $r=0.70,0.70$ and $0.79$,
between $t$ and $CO$, shows up for chain lengths $N=54,64$ and $80$
respectively.}
\end{enumerate}

\section{DISCUSSION AND CONCLUSIONS}
The 'turning point' value of $CO=0.17$ is
actually the peak of the homopolymer relaxation histogram 
distribution as previously discussed.
This means that $CO$ and folding time are positively correlated only 
for proteins with predominantly non-local contacts.
We interpret this result as a consequence of the properties of the move set
used to explore the conformational space together with the ruggedness of
the energy landscape. As seen in section~\ref{sec:3a}, kink-jump dynamics does
not favour the formation of high $CO$ structures in homopolymers.
In proteins, when the native structure is of high $CO$, it will be difficult to
escape from kinetic traps associated with local energy minima and
structures of lower $CO$. This confirms and explains our
previous findings {\cite{PFN}} according to which the folding
performance achievable is strongly sensitive to target conformation for
chain lengths $N \geq 80$.\par
The comparison of the simulation's results
with the experimental data set of 24 two-state proteins,
with chain length ranging from 41 to 154 amino acids, reported
in Ref.3 is hindered by the fact that the proteins considered
in Ref.3 fail to exhibit the scaling of folding time with 
chain length which is typical of lattice model simulations.
However, a strong correlation ($r=0.80$) is also found between $CO$ 
and the folding times. Moreover, this correlation is considerably
improved ($r=0.97$) if only long protein chains ($N \geq 80$) are considered. 
\par  
As a general conclusion, we might say that results on lattice
models encourage the idea that the contact order of the native structure
plays a significant role in determining the folding rate. The match with the 
correlation between the $CO$ and the folding time found from the analysis of
experimental data suggests that lattice polymer dynamics with local moves
does capture the key dynamical features of real protein folding.\par 
It would be interesting to know if similar results can be found
in the scope  of 'off-lattice' models, where one would expect proteins with
high helical content to be better folders.

\acknowledgments{P.F.N.F would like to thank Dr.
A. Nunes for helpful suggestions, and Programa Praxis XXI for financial support.}

\begin{figure*}[!h]
\psfig{file=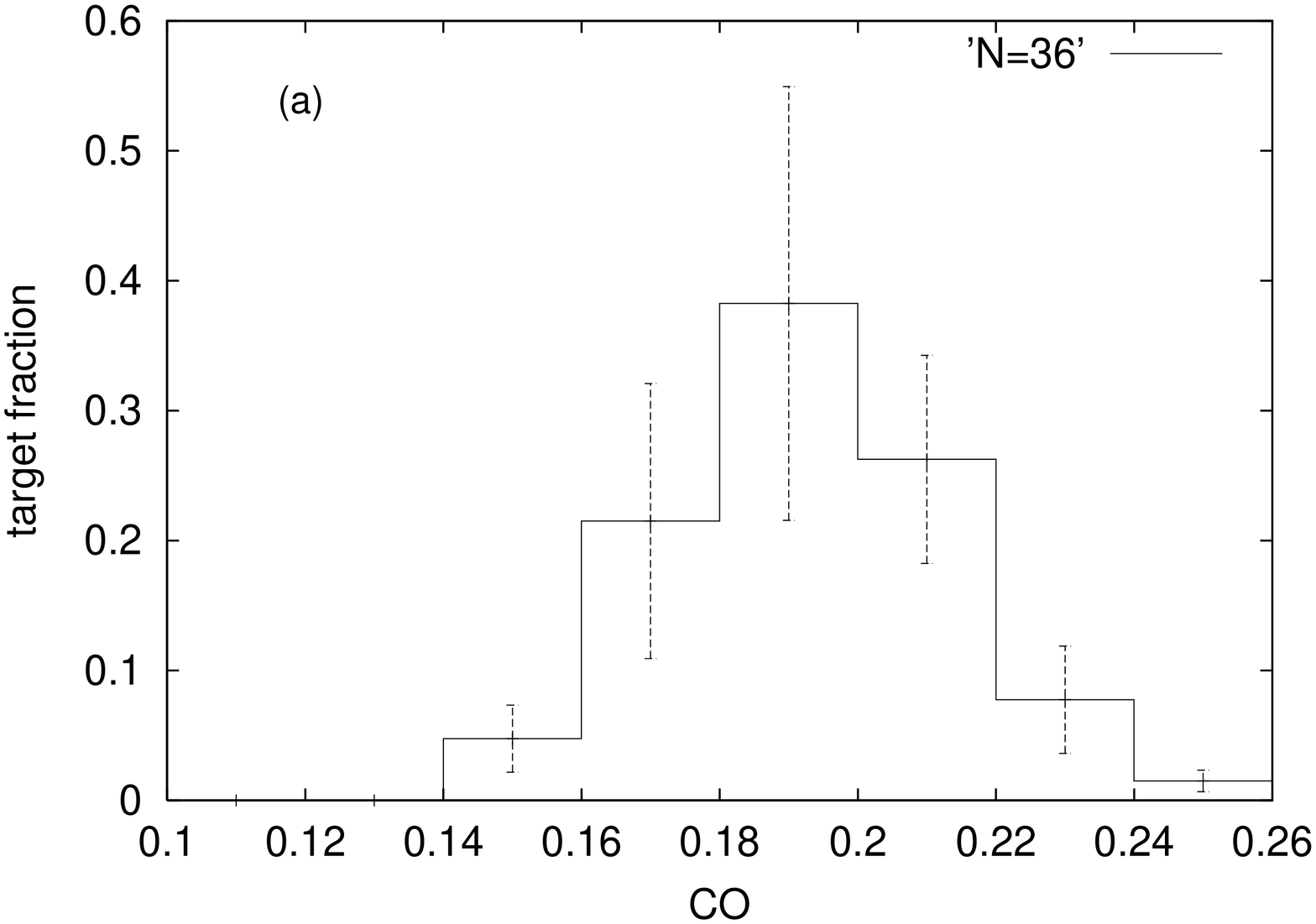, height=7cm, width=7cm,clip=,angle=270}
\psfig{file=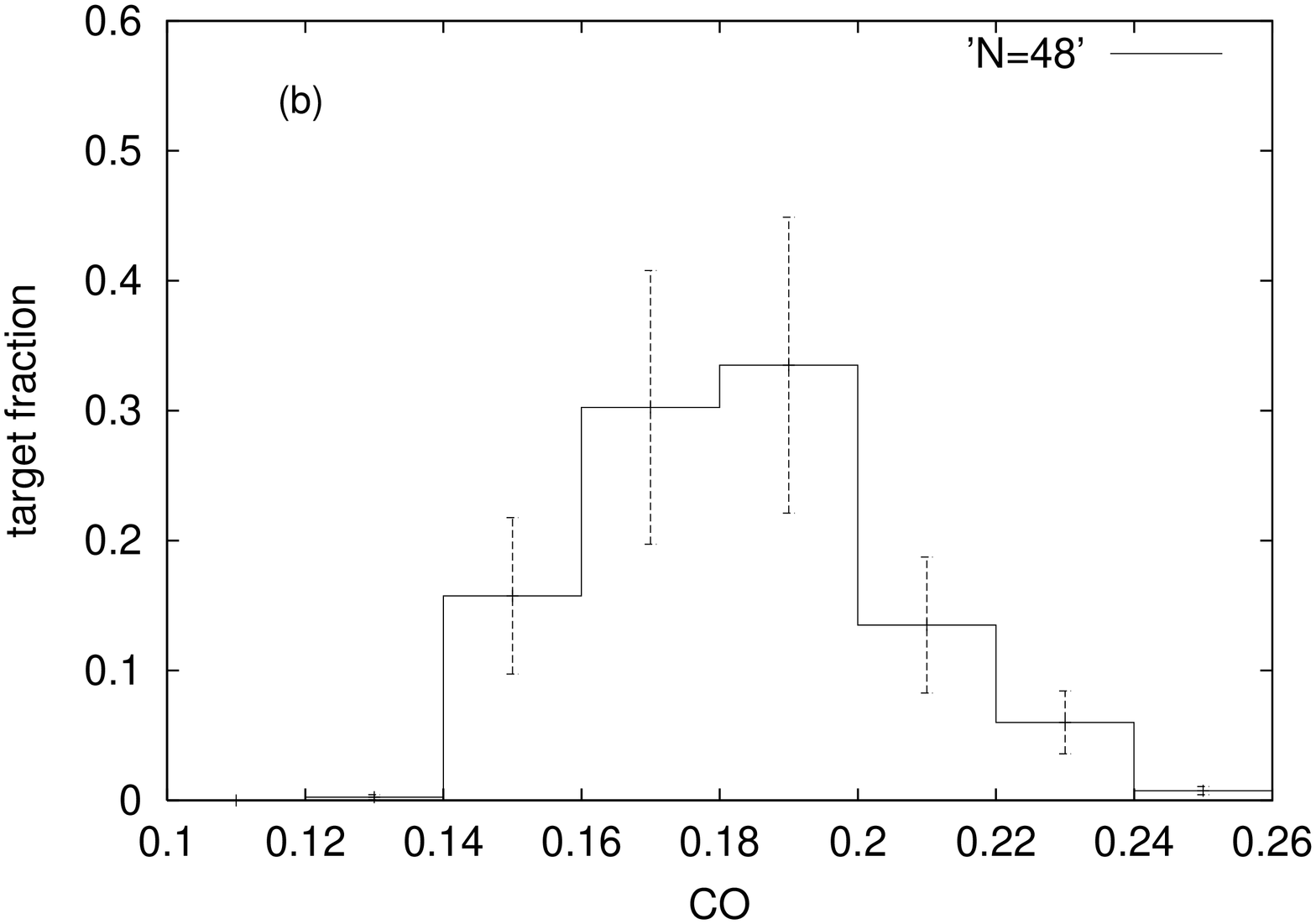, height=7cm, width=7cm,clip=,angle=270}\\
\psfig{file=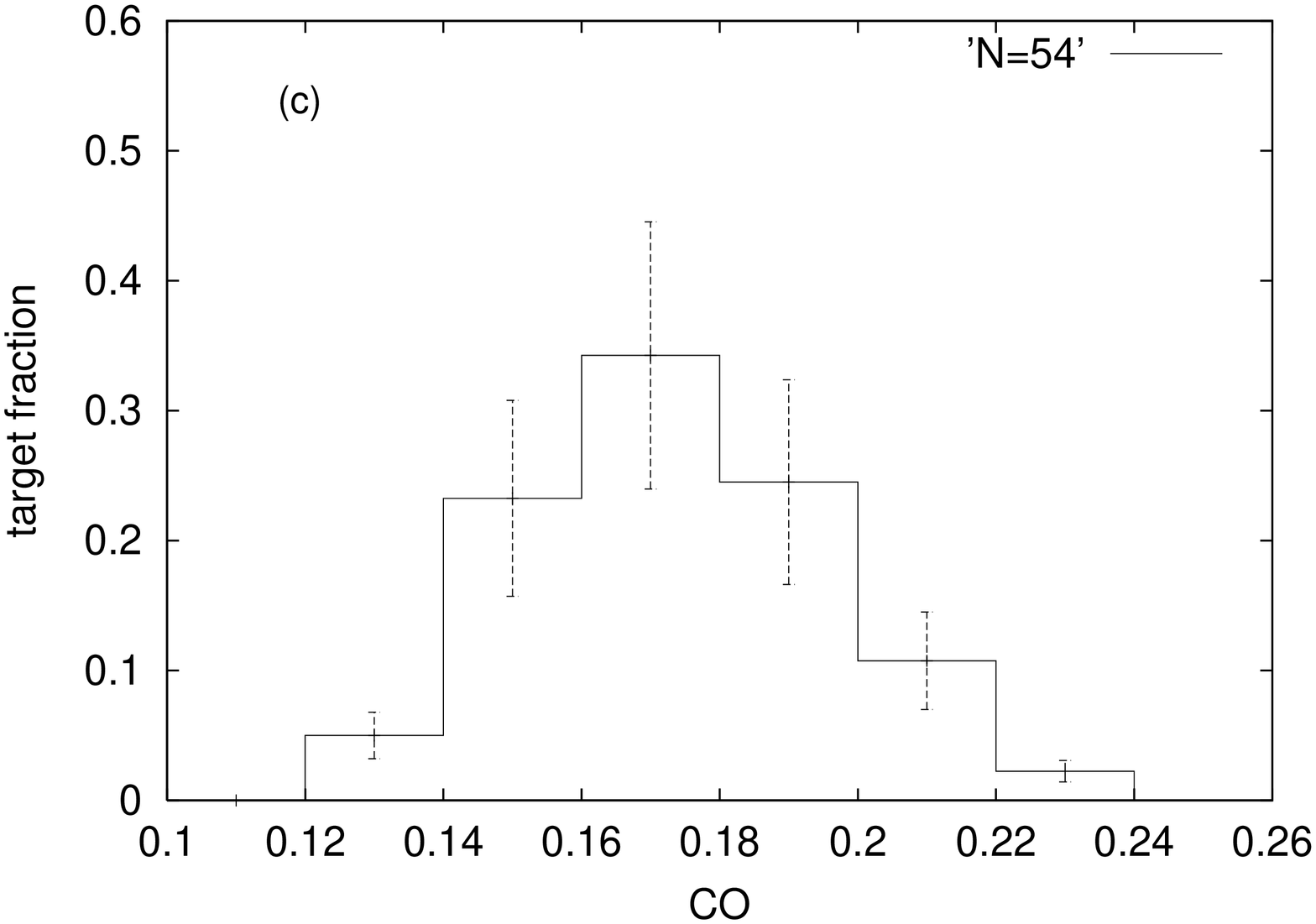, height=7cm, width=7cm,clip=,angle=270}
\psfig{file=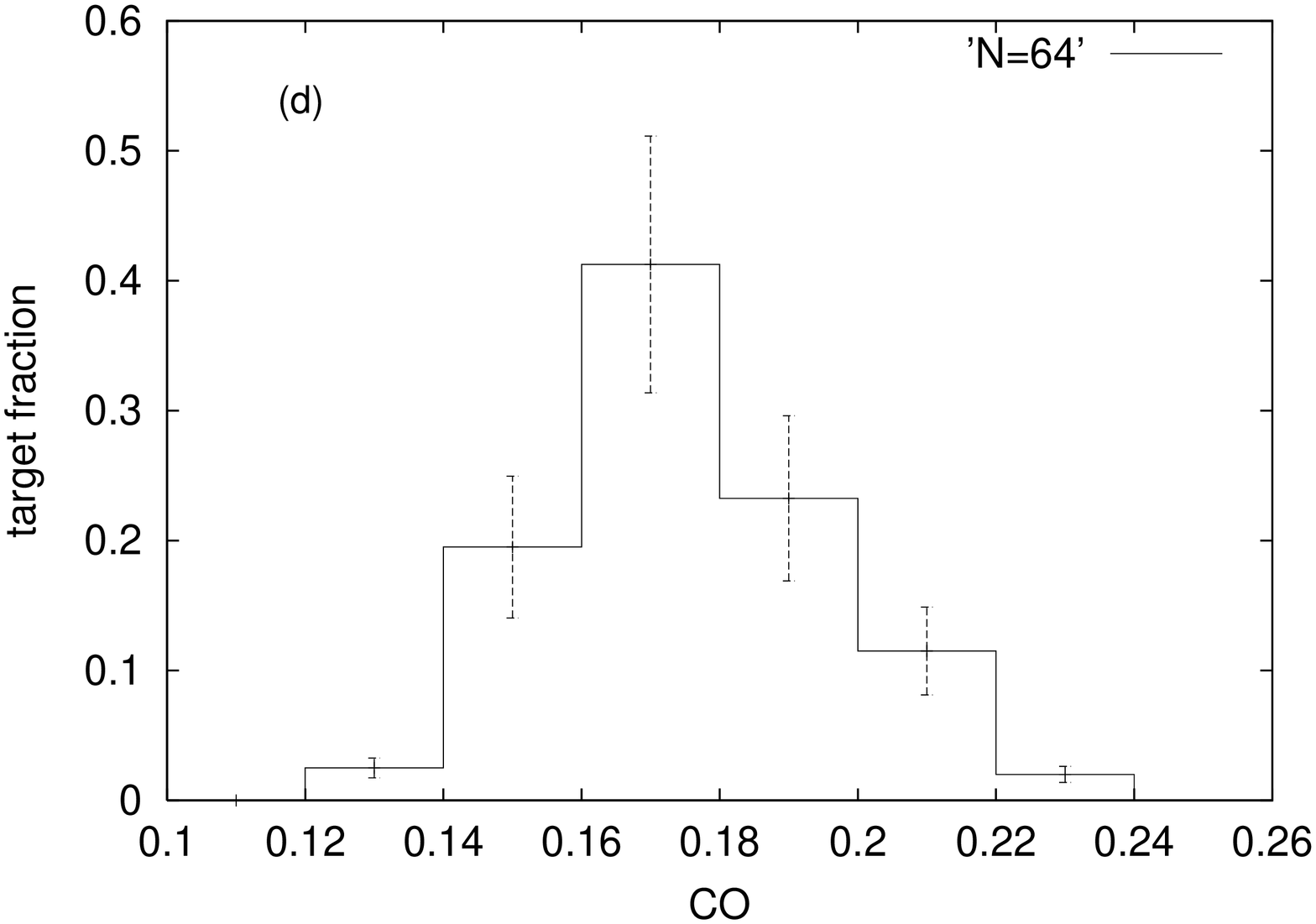, height=7cm, width=7cm,clip=,angle=270}\\
\psfig{file=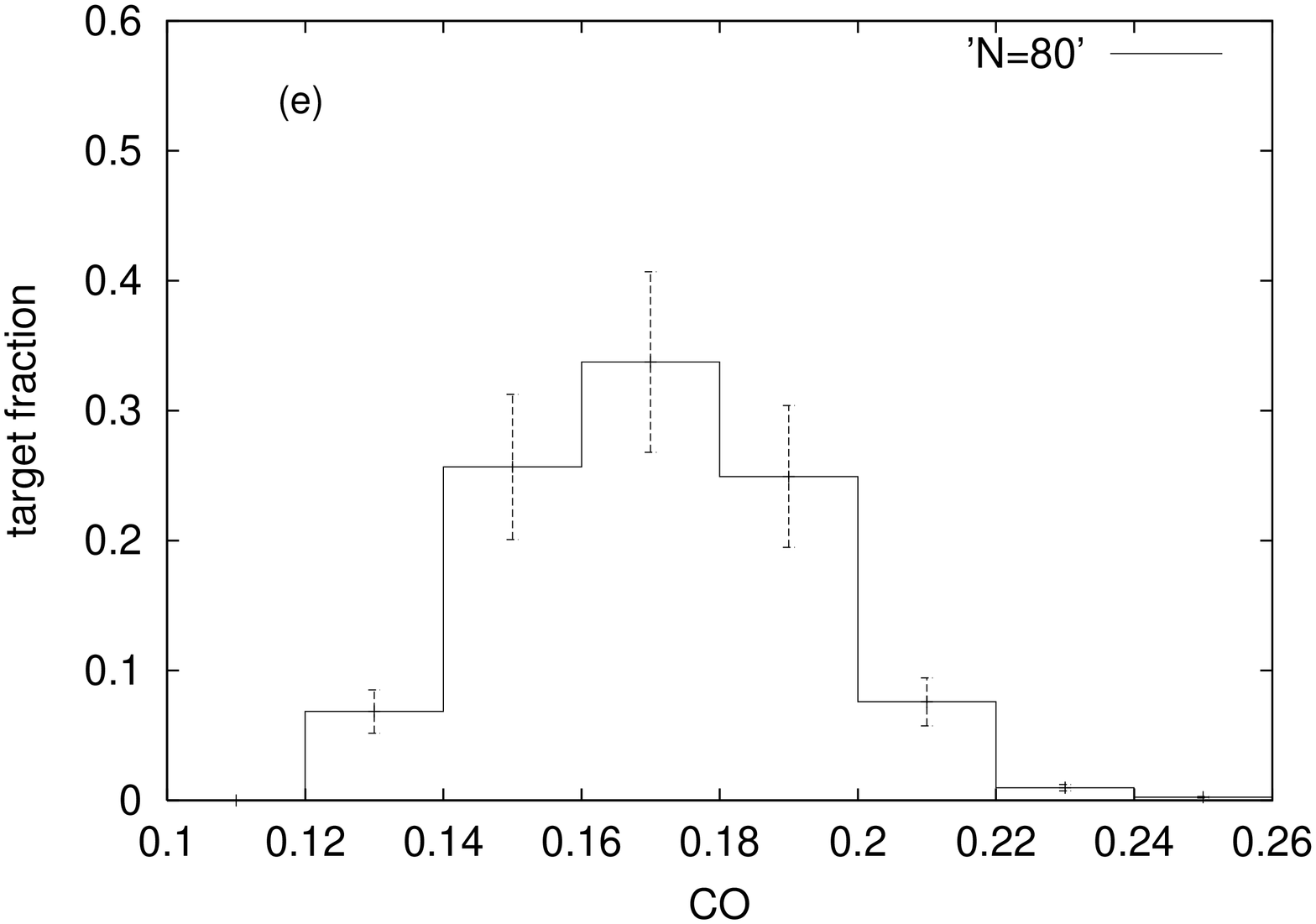, height=7cm, width=7cm,clip=,angle=270}
\psfig{file=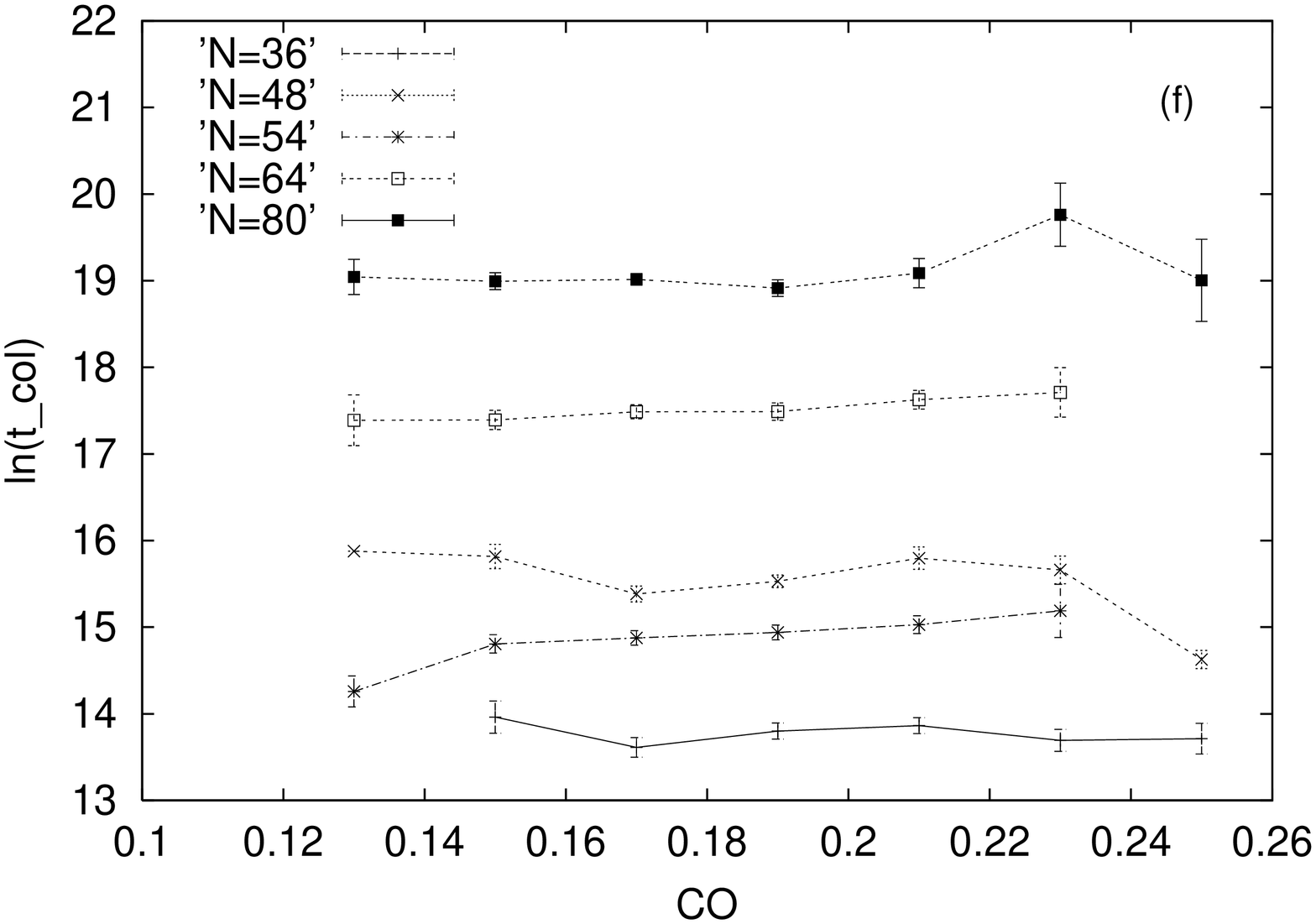, height=7cm, width=7cm,clip=,angle=270}
\caption{(a)-(e) Histograms of the distribution of the relative contact order parameter;
for each studied chain length, $N$, a sample of 400 targets was considered.
(f) Shows the dependence of the collapsing time, $t_{col}$, 
on $CO$. $t_{col}$ was computed as the mean first
passage time (FPT) averaged over 400 simulation runs.}
\label{figure:no1}
\end{figure*}

\begin{figure}
\psfig{file=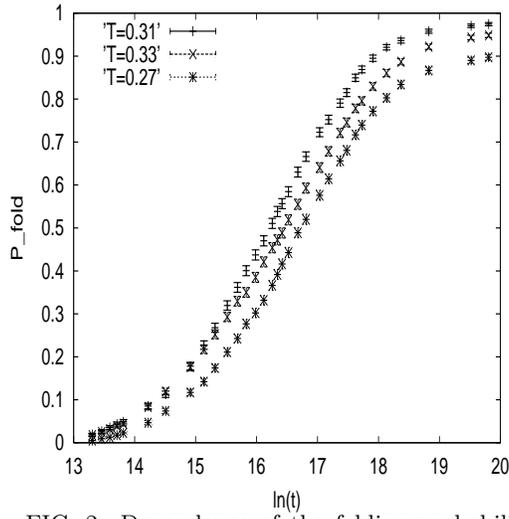, height=7cm, width=7cm, angle=270}
\caption{Dependence of the folding probability, $P_{fold}$, on  $log(t)$ at three different temperatures. $P_{fold}$ was computed as the number of folding simulations which ended up to time $t$
normalised to the total number of attempted runs.
For each curve $\approx$2000 simulations were used distributed acroos the available 48 bead long targets and sequences. }
\label{figure:no2}
\end{figure}

\begin{figure}
\psfig{file=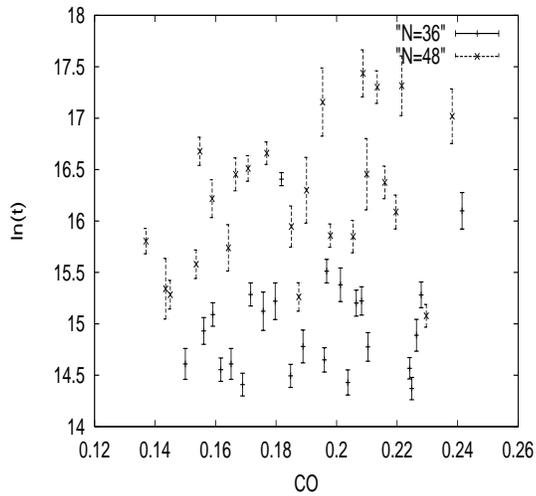, height=7cm, width=7cm, angle=270}
\caption{Dependence of the folding time, $t$, on the relative contact order, $CO$, for 36 and 48 bead long targets.}
\label{figure:no3}
\end{figure}

\begin{figure*}
\psfig{file=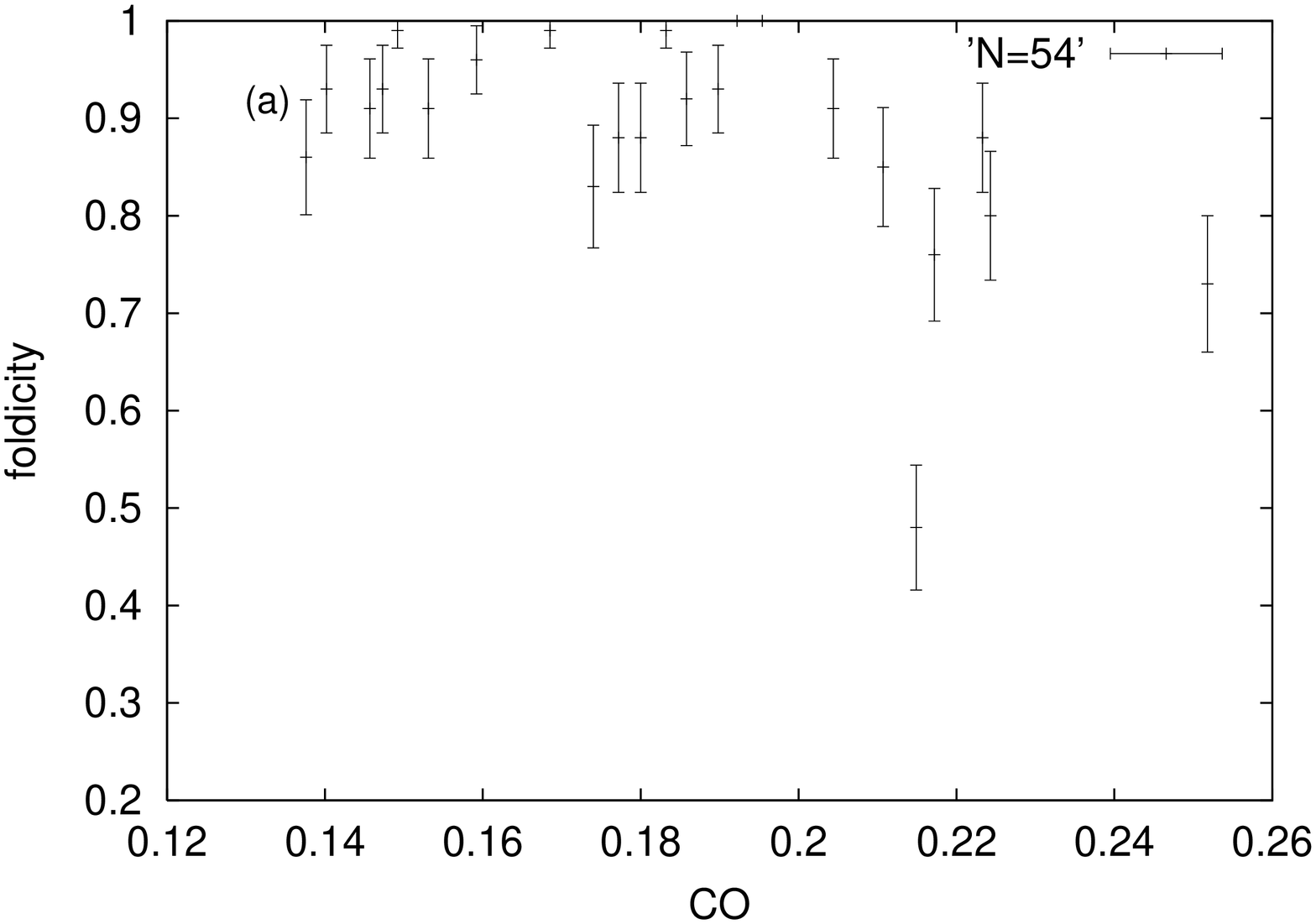, height=7cm, width=7cm,clip=,
angle=270}
\psfig{file=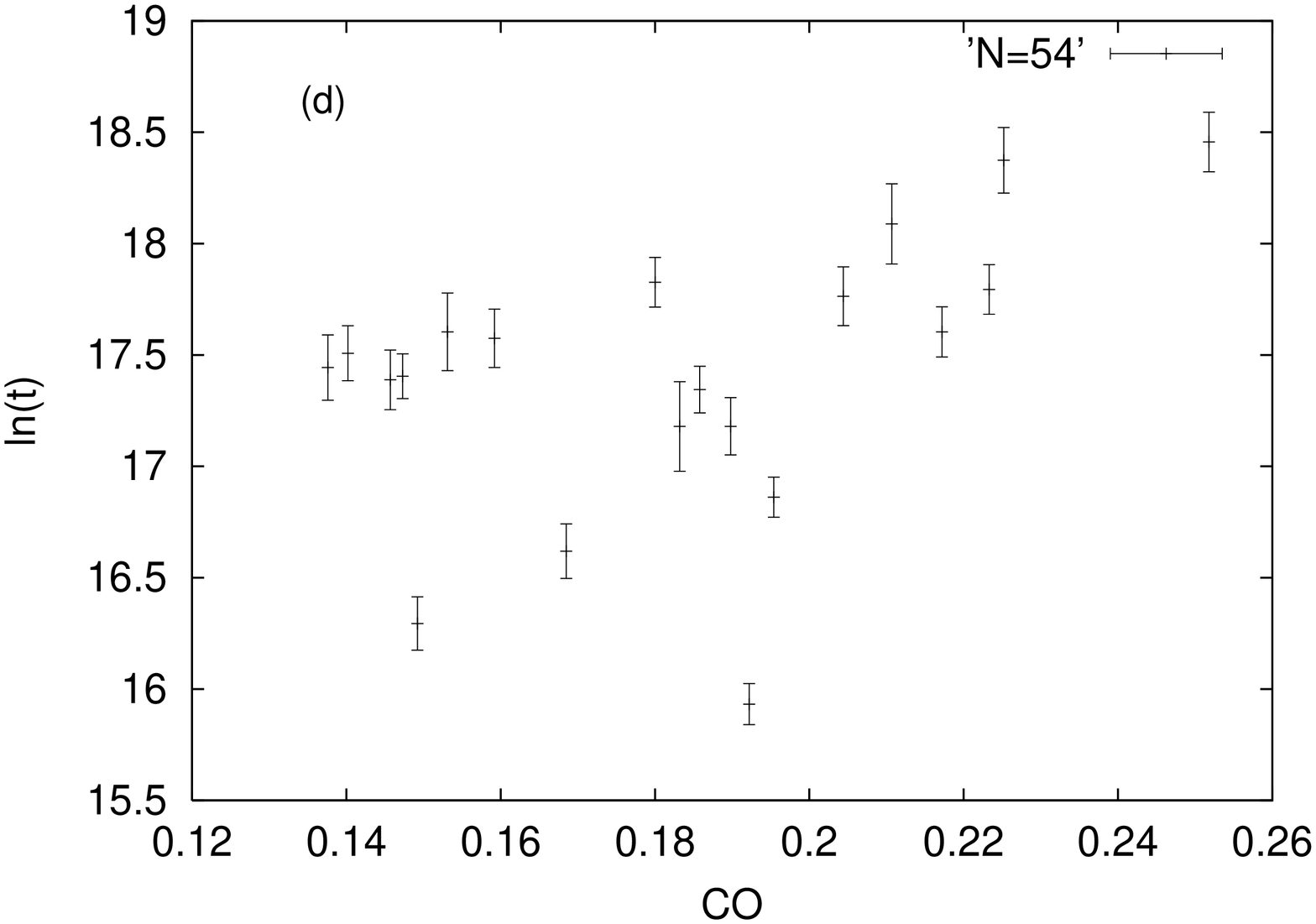, height=7cm, width=7cm,clip=, angle=270}
\end{figure*}
\begin{figure*}
\psfig{file=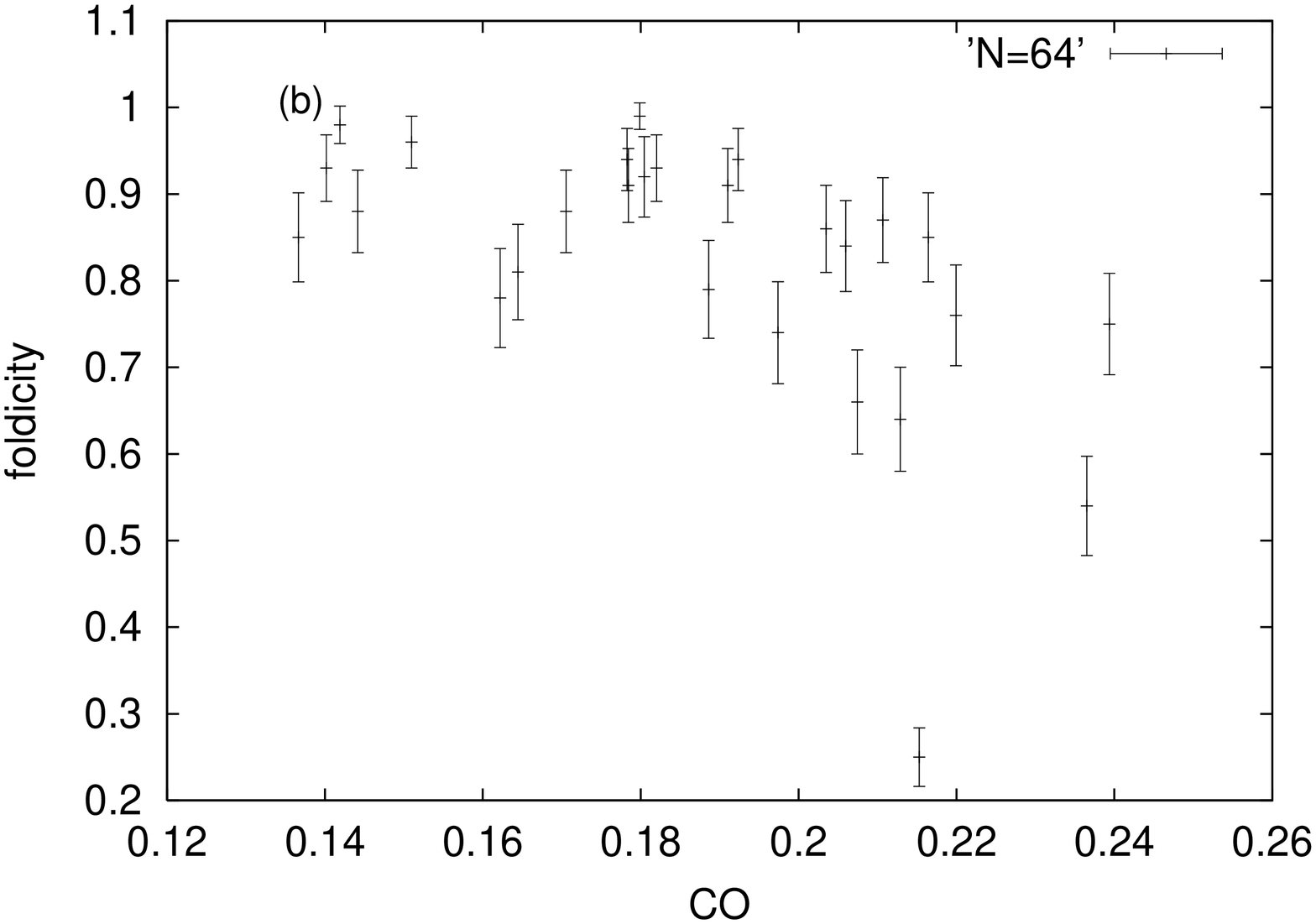, height=7cm, width=7cm,clip=,
angle=270}
\psfig{file=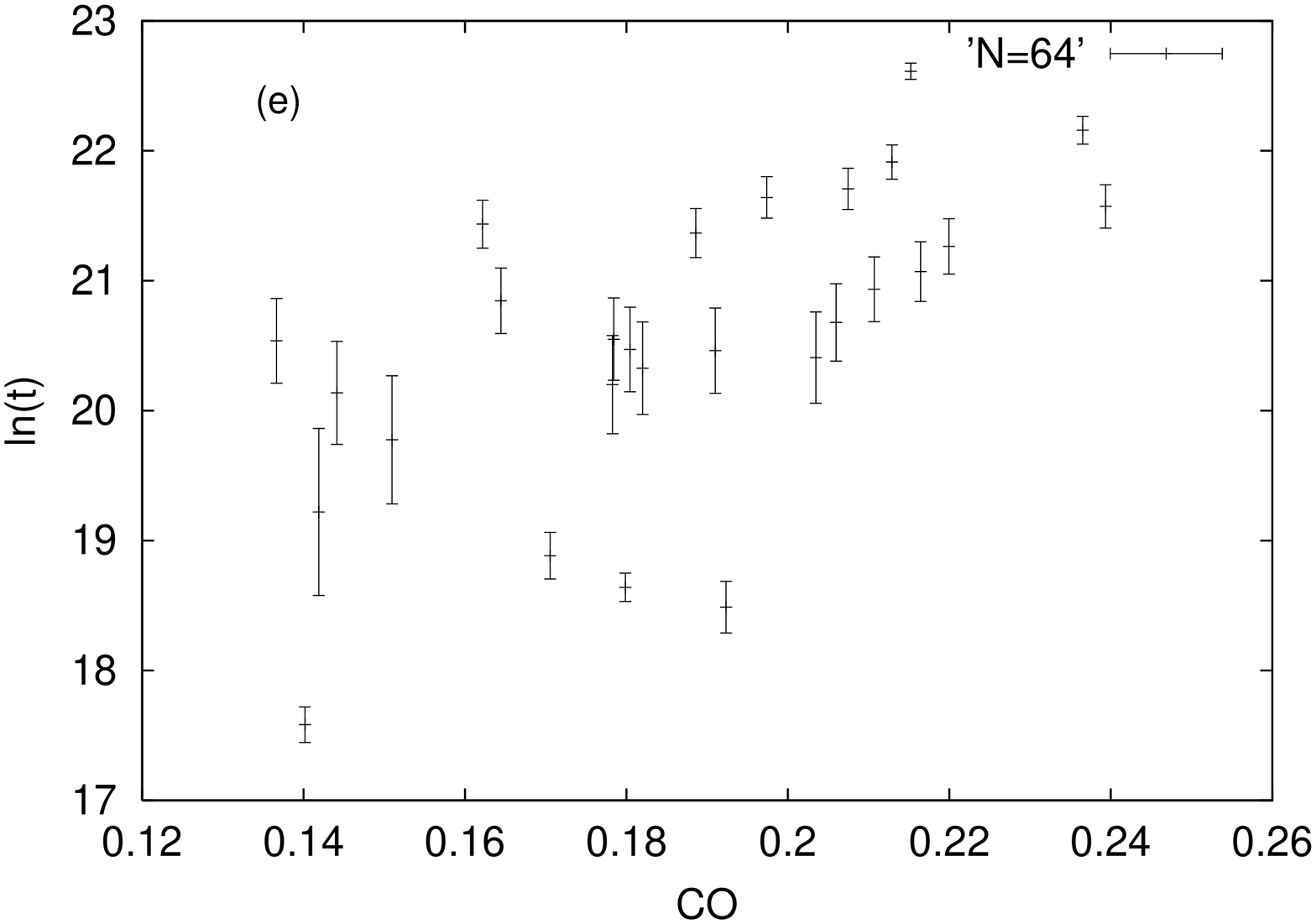, height=7cm, width=7cm,clip=, angle=270}
\end{figure*}
\begin{figure*}
\psfig{file=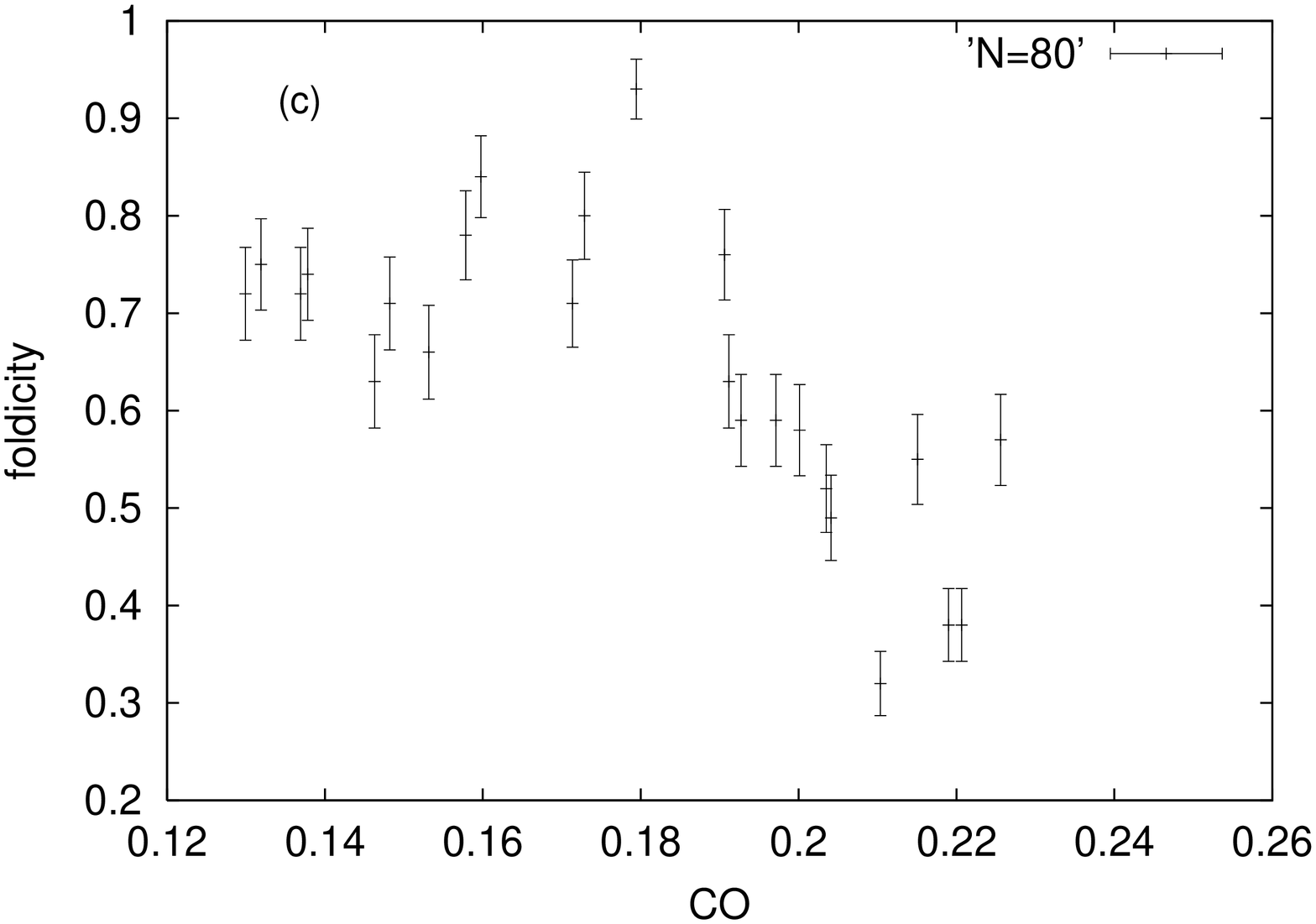, height=7cm, width=7cm,clip=,
angle=270}
\psfig{file=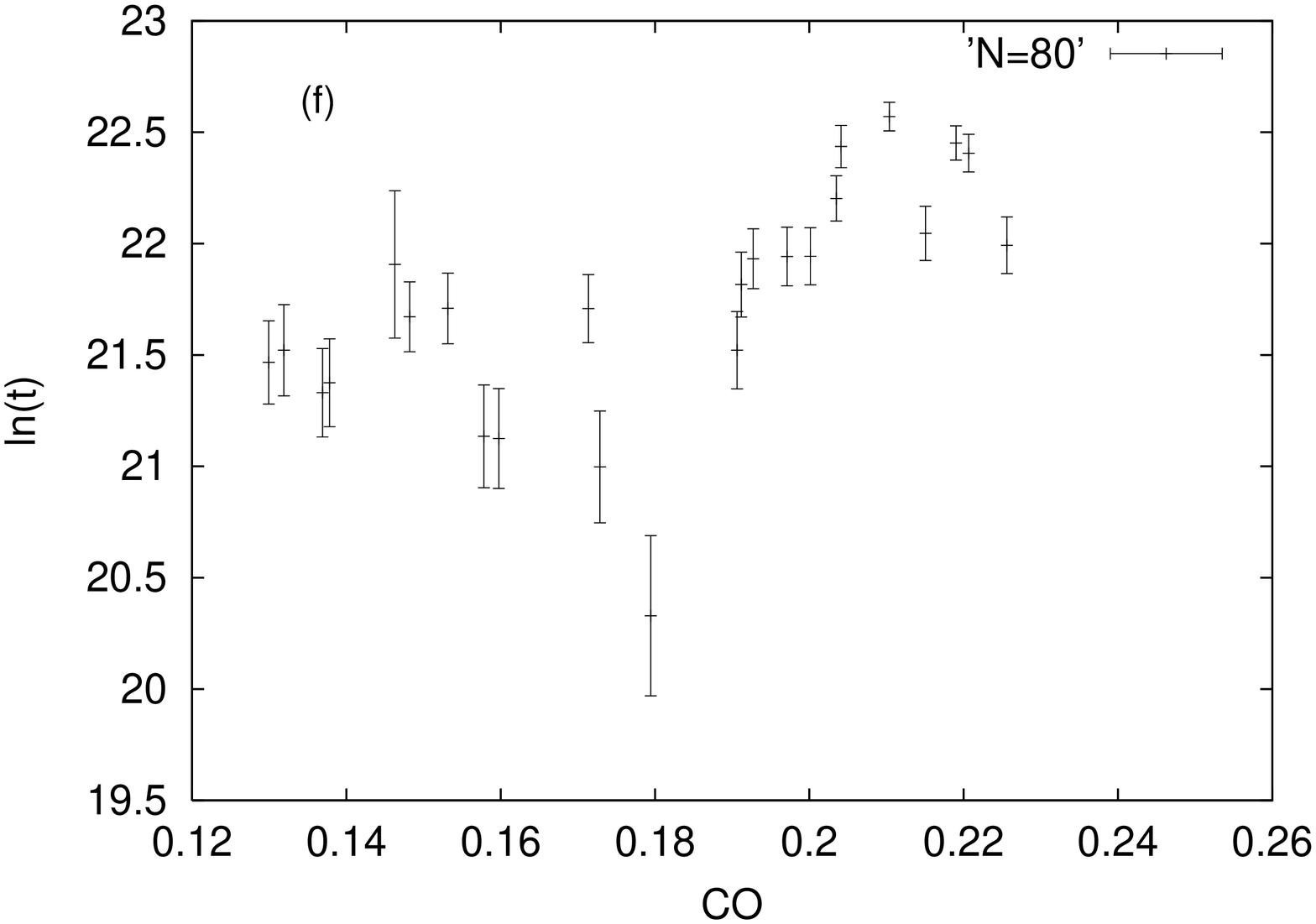, height=7cm, width=7cm,clip=, angle=270}
\caption{(a)-(c)Dependence of foldicity on the relative contact  
order, $CO$, for 54, 64 and 80 bead long targets. (d)-(f) show
the dependence of the (estimated) folding time, $t$, on $CO$.}
\label{figure:no4}
\end{figure*}

\begin{thebibliography}{99}
\bibitem{ALLAN}{A.R. Fersht, Proc. Natl. Acad. Sci. USA {\bf 97}, 1525 (2000)}
\bibitem{PLAXCO1}{K.W. Plaxco, K.T. Simons and D. Baker, J. Mol. Biol. {\bf 277}, 985 (1998)}
\bibitem{PLAXCO}{K.W. Plaxco, K.T. Simons, I. Ruczinski and D. Baker, Biochemistry {\bf 39},
11177 (2000)}
\bibitem{THOMAS}{T.M. Fink and R.C. Ball, Physica D {\bf 107}, 199 (1997)}
\bibitem{PFN}{P.F.N. Faisca and  R.C. Ball, J. Chem. Phys. {\bf 116}, 7231 (2002)}
\bibitem{S1}{A.M. Gutin, V.I. Abkevich and E.I. Shakhnovich, Phys. Rev. Lett. {\bf 77},
5433 (1996)}
\bibitem{DIMITRIEVSKI}{K. Dimitrievski, B. Kasemo and V.P. Zhdanov, J. Chem. Phys. {\bf 113},
883 (2000)}
\bibitem{DEMCHENKO}{A.P. Demchenko, Curr. Prot. and Peptide Sci {\bf 2}, 73 (2001)}
\bibitem{ABKEVICH}{V.I. Abkevich, A.M. Gutin and E.I. Shakhnovich, J. Mol. Biol. {\bf 252},
460 (1995)}
\bibitem{Wetlaufer}{D.B. Wetlaufer, Proc. Natl. Acad. Sci. USA {\bf 70}, 697 (1973)}
\bibitem{BAKER}{D. Baker, Nature {\bf 405}, 39 (2000)}
\bibitem{LINDBERG}{M.O. Lindberg, J. Tangrot, D.E. Otzen, D.A.
Dolgikh, A.V. Finkesltein and M. Oliveberg, J. Mol. Biol. {\bf 314}, 891 (2001)}
\bibitem{MJ}{S. Miyazawa and R. Jerningan, Macromolecules {\bf 18}, 534 (1985)}
\bibitem{METROPOLIS}{N. Metropolis, A. Rosenbluth, M.N. Rosenbluth and A.H. Teller, 
J. Chem. Phys. {\bf 21}, 1087 (1958)}
\bibitem{BINDER}{D.P. Landau and K. Binder, 
{\it A Guide to Monte Carlo Simulations in Statistical Physics}, 
Cambridge University Press (2000)}
\bibitem{GUTIN}{V.I. Abkevich, A.M. Gutin and E.I. Shakhnovich, J. Mol. Biol. {\bf 252},
460 (1995)}
\bibitem{SG}{E.I. Shakhnovich, A.M. Gutin, Proc. Natl. Acad. Sci. USA {\bf 90}, 7195 (1993)}
\end{thebibliography}
\end{document}